\documentstyle[amssymb,prb,twocolumn,aps,epsfig]{revtex} 
 
\begin{document} 
\draft 
 
\twocolumn[\hsize\textwidth\columnwidth\hsize\csname 
@twocolumnfalse\endcsname 
 
\title{Kondo resonances and Fano antiresonances in transport 
through quantum dots} 
\author{M. E. Torio$^1$, K. Hallberg$^2$, A. H. Ceccatto$^1$, and C. R. Proetto$^2$} 
\address{$^1$Instituto de F\'{\i}sica de Rosario, CONICET-UNR, Bv. 27 de Febrero 210bis, 2000 
Rosario, Argentina} 
\address{$^2$Centro At\'{o}mico Bariloche and Instituto Balseiro, 
Comisi\'{o}n Nacional de Energ\'{\i}a At\'{o}mica 
8400 Bariloche, Argentina} 
 
\maketitle 
 
\begin{abstract} 
The transmission of electrons through a non-interacting tight-binding chain 
with an interacting {\it side} quantum dot (QD) is analized. When the Kondo effect develops at 
the dot the conductance presents a wide minimum, reaching zero at the unitary limit. 
This result is compared to the opposite behaviour found in an embedded QD. 
Application of a magnetic field destroys the Kondo effect and the conductance 
shows pairs of dips separated by the charging energy $U$. 
The results are discussed in terms of Fano antiresonances and explain qualitatively recent 
experimental results. 
\end{abstract} 
 
\pacs{PACS numbers: 73.63.-b, 72.15.Qm} 
\vskip2pc] 
 
\narrowtext

Semiconductor quantum dots (QD) are small droplets of electrons, confined in 
the three spatial directions. Energy and charge quantization results from 
this confinement. As both features are shared with real atomic systems, from 
the very beginning an extremely useful analogy has been exploited between ``$%
\mathop{\rm real}%
$'' and ``artificial'' atomic systems. This analogy received strong support 
through an experimental breakthrough where the Kondo effect in quantum dots 
was unambiguously measured.\cite{david,sara} Historically, the Kondo effect 
was introduced about forty years ago to explain the resistivity minimum for 
decreasing temperatures observed in metallic matrices with a minute fraction 
of magnetic impurities.\cite{jun} According to the detailed microscopic 
theory, when the temperature $T$ decreases below the Kondo temperature $T_{K} 
$, the localized magnetic impurity starts to interact strongly with the 
surrounding electronic cloud, which finally results in a singlet many-body 
ground state, reaching its maximum strength at $T=0.$\cite{hewson} The 
minimum in the resistivity results from the fact that, as the temperature is 
lowered, the scattering with phonons decreases down to the Kondo temperature 
at which the scattering with localized impurities becomes important as the 
Kondo effect is operative. It is important to emphasize that in this case, 
the so-called traditional Kondo effect, magnetic impurities act as 
scattering centers, {\it increasing} the sample resistivity. 
 
The opposite behavior is found in the Kondo effect in quantum dots. The 
situation considered almost without exception both theoretically and 
experimentally, consists of a quantum dot connected to two leads in such a 
way that electrons transmitted from one electrode to other should 
necessarily pass through the quantum dot (a {\it substitutional} dot). As 
theoretical calculations predicted,\cite{glazman,patrick} in this 
configuration the conductance {\it increases} when decreasing the 
temperature and the Kondo effect sets in, essentially due to a resonant 
transmission through the so-called Kondo resonance which appears in the 
local density of states at the dot site at the Fermi level. In this situation, at $T=0$, 
the conductance should take the limiting value $2e^{2}/h,$ corresponding to the 
unitary limit of a one-dimensional perfect transmission channel.\cite{leo} 
 
The aim of this work is to analyze an alternative configuration of a {\it %
side-coupled} quantum dot, attached to a perfect quantum wire. In 
this case, the quantum dot acts as a scattering center for 
transmission through the chain, in close analogy with the 
traditional Kondo effect. We have found that when the dot provides 
a resonant energy for scattering, the conductance has a sharp 
decrease, reminiscent of the Fano antiresonances observed in 
scanning tunneling microscope experiments for magnetic atoms on a 
metallic substrate.\cite{li} A similar problem has 
been discussed both by Kang {\it et al.}\cite{kang} and by Bulka 
{\it et al.}\cite{bulka}. However, in \cite{kang} the authors use 
the approximate slave boson mean field theory to describe the 
Kondo regime in the limit $U\rightarrow \infty$, while in 
\cite{bulka} the configuration considered is different from ours.  In this
paper we show that by using a very precise  numerical technique appropriate
for this system we can incorporate  the effects of a realistic value of the
charging energy $U$. This  allows us to obtain the features observed
experimentally like the  double dip structures of Fig. 3, which give rise to
the  diamond-shaped features in differential conductance experiments. 
\cite{gores} \\ 
 
\begin{figure}[tbp] 
\vspace{-.5cm} 
\epsfxsize=8cm 
\epsfysize=3cm 
\centerline{\epsfbox{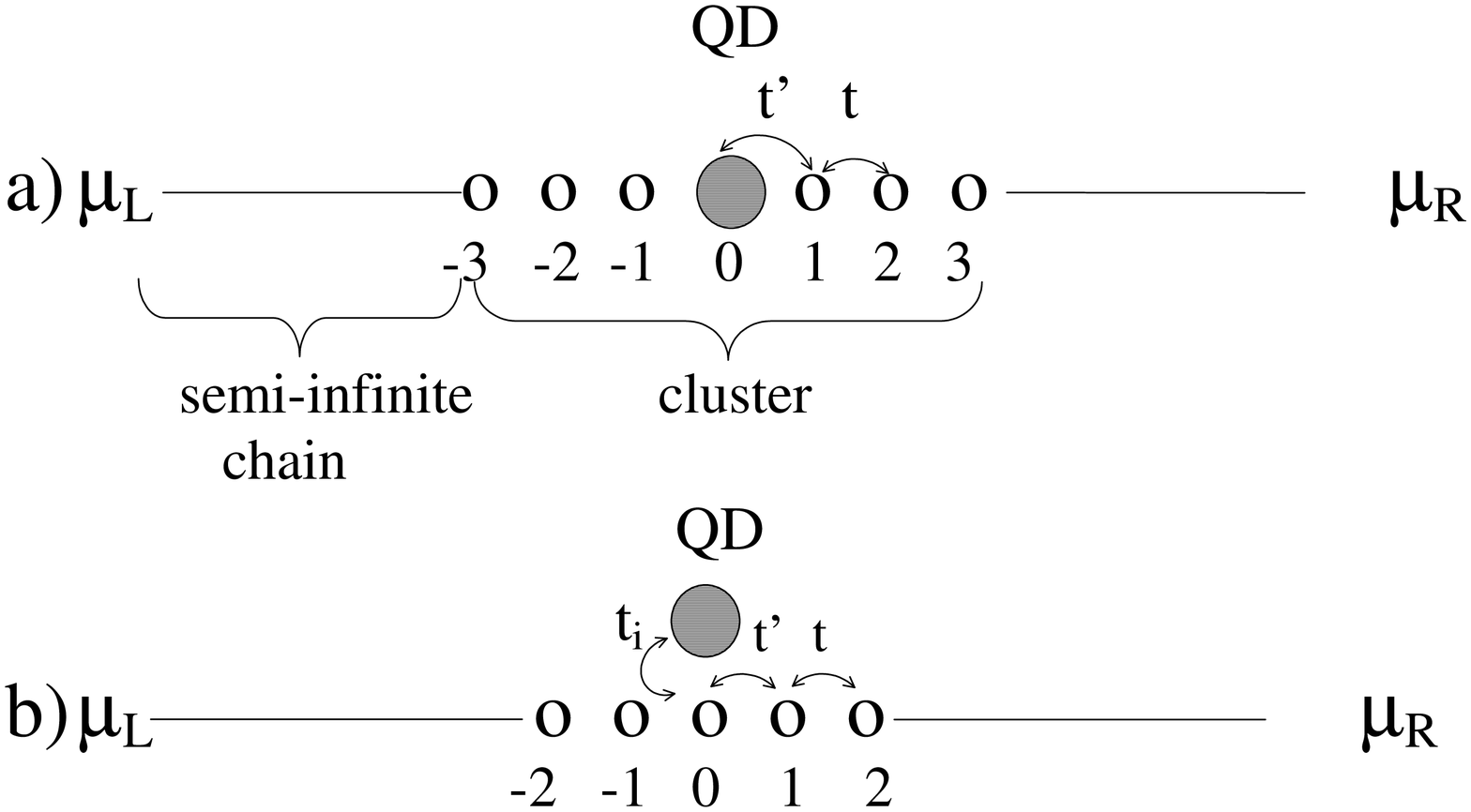}} 
\narrowtext 
\caption{Schematic representation of our models: a) {\it substitutional} dot 
configuration; b) {\it side} dot configuration. Small open circles represent 
cluster non-interacting sites, big full circle represent the dot site. The 
cluster includes the dot and a few non-interacting sites. Left and right 
full lines represent the non-interacting tight-binding semiinfinite chains with 
their respective chemical potentials $\mu_L$ and $\mu_R$.} 
\end{figure} 
 
The models employed in the calculation are schematically shown in Fig. 1. 
Case a), corresponding to the {\it substitutional} dot situation, consists 
of two semiinfinite non-interacting tight-binding chains connected to a 
central site (the dot). Case b), corresponding to the {\it side} dot, 
consists of a quantum wire coupled sideways to a QD. The dot is modeled as 
an Anderson impurity. The Hamiltonian reads: 
 
\begin{equation} 
{\cal H}={\cal H}_{0}+{\cal H}_{int}, 
\end{equation} 
where ${\cal H}_{0}$ is the Hamiltonian of two semiinfinite chains, 
\begin{eqnarray} 
{\cal H}_{0}=&-&\sum_{j\leq -1,\sigma }t_{j}(c_{j\sigma }^{\dagger }c_{j+1\sigma }+%
\text{h.c.})  \nonumber \\ 
&-&\sum_{j\geq 0,\sigma }t_{j}(c_{j\sigma }^{\dagger }c_{j+1\sigma }+\text{%
h.c.}),  \eqnum{2a} 
\end{eqnarray} 
and ${\cal H}_{int}$ can be written as 
\begin{equation} 
{\cal H}_{int}^{a}=\sum_{\sigma }\varepsilon _{\sigma }\text{ }c_{0\sigma 
}^{\dagger }c_{0\sigma }+\frac{U}{2}\text{ }n_{0\sigma }n_{0\overline{\sigma 
}}  \eqnum{2b} 
\end{equation} 
for the {\it substitutional} dot configuration, while 
\begin{equation} 
{\cal H}_{int}^{b}=\sum_{\sigma }-t_{i}(c_{0\sigma }^{\dagger }c_{\sigma }+\text{%
h.c.})+\varepsilon _{\sigma }\text{ }c_{\sigma }^{\dagger }c_{\sigma }+\frac{%
U}{2}\text{ }n_{\sigma }n_{\overline{\sigma }}  \eqnum{2c} 
\end{equation} 
for the {\it side} dot configuration. In the equations above, $t_{j}=t$ for $%
j\eqslantless -2$, $j\geqslant 1,$ while $t_{-1}=t_{0}=t^{\prime }$, $n_{0\sigma 
}=c_{0\sigma }^{\dagger }c_{0\sigma }$, $n_{\sigma }=c_{\sigma }^{\dagger 
}c_{\sigma }$ and $U>0$. As we are also interested in the behavior in a magnetic 
field $H$, we consider the local energies as $\varepsilon _{\uparrow 
}=\varepsilon _{0}+\Delta \varepsilon /2,$ $\varepsilon _{\downarrow 
}=\varepsilon _{0}-\Delta \varepsilon /2,$ with $\Delta \varepsilon =g\mu 
_{B}H$ the Zeeman splitting of the localized orbital, {\it i. e. } the 
principal magnetic field effect is to shift the local QD levels.\cite{meir2} 
It is interesting to point out that both models could be mapped to a single 
semiinfinite chain, with the dot sitting at the free end, and the remaining 
sites corresponding to the even basis states that couples to the dot.\cite 
{jorge} Both models become exactly equivalent from the point of view of 
their equilibrium properties,\cite{schlott} if the hoppings are related as 
follows: $t_{i}=\sqrt{2}t^{\prime }=t.$ However, as we show below, the 
transport properties of both models are completely different. 
 
For the analysis of our transport results we have used the following 
equation for the magnetic field dependent conductance,\cite{meir} 
in the linear response regime ($\mu_L\to 0^+$ and $\mu_R\to 0^-$) 
\begin{equation} 
G(H)=\frac{e^{2}}{h}\frac{2\pi t^{\prime }{}^{2}}{t}\sum_{\sigma }\rho 
_{\sigma }(\omega =0)  \eqnum{3} 
\end{equation} 
where $\rho _{\sigma }(\omega )$ is the local density of states (per spin) 
at site $0$ evaluated at the Fermi energy $\omega =0.$ 
 
To obtain the density of states $\rho _{\sigma }(\omega =0)$ we use a 
combined method. In the first place we consider an open finite cluster of $N$ 
sites ($N=7$ for case (a) and 6 for case (b)) which includes the impurity. This is diagonalized 
using the exact diagonalization Lanczos technique\cite{lanczos}. We then 
proceed to embed the cluster in an external reservoir of electrons, which 
fixes the Fermi level of the system, attaching two semiinfinite leads to its 
right and left\cite{ferrari}. This is done by calculating the one-particle 
Green's function $\hat{G}$ of the whole system within the chain approximation 
of a cumulant expansion\cite{cumulant} for the dressed propagators. This 
leads to the Dyson equation $\hat{G}=\hat{G}\hat{g}+\hat{T}\hat{G}$, where $%
\hat{g}$ is the cluster Green's function obtained by the Lanczos method. 
Following Ref. 16, the charge fluctuation inside the cluster is taken into 
account by writing $\hat{g}$ as a combination of $n$ and $n+1$ particles 
with weights $1-p$ and $p$ respectively: $\hat{g}=(1-p)\hat{g}_{n}+p\hat{g}%
_{n+1}$. The total charge of the cluster and $p$ are calculated by solving 
selfconsistently the equations: 
\begin{eqnarray} 
Q_{c} &=&n(1-p)+(n+1)p,  \eqnum{4} \\ 
Q_{c} &=&-\frac{1}{\pi }\int_{-\infty }^{\epsilon _{F}}\sum_{i}\text{Im }%
G_{ii}(\omega )\text{ }d\omega ,  \eqnum{5} 
\end{eqnarray} 
where $i$ runs on the cluster sites. Once convergence is reached, 
the density of states is obtained from $\hat{G}$. It 
is important to stress that this method is reliable only if 
$t^{\prime }$ is large enough, so that the Kondo cloud is about the 
size of the cluster. Moreover, the fact that the 
conductance reaches the unitary limit for the symmetric case 
provides a crucial test of the validity of the method (see 
below).\\ 
\vspace{-1.cm}
\begin{figure}[tbp] 
\epsfxsize=7.5cm 
\epsfysize=8cm 
\centerline{\epsfbox{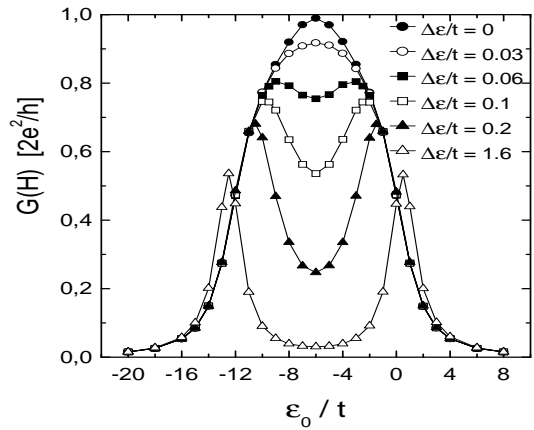}} 
\narrowtext
\vspace{-1.5cm}
\caption{Conductance (in units of $2e^{2}/h$) for the {\it substitutional} dot 
configuration, as a function of impurity level position and for several values 
of the magnetic field ($t^{\prime}/t=1/\sqrt{2}$ and $U/t=12$)}. 
\end{figure} 
 
In order to compare both geometrical realizations we present in Fig. 2 
the conductance  for the {\it substitutional} QD (Fig. 1a). As discussed above, at 
zero-temperature and magnetic field, the Kondo resonance, which develops 
right at the Fermi level, greatly enhances the transmission when the average 
dot occupancy is close to (but less than) 1, {\it i.e} in the Kondo regime. For the 
symmetric situation $\varepsilon _{0}=-U/2$, the transmission is perfect and 
the conductance reaches the unitary limit $G(0)=2e^{2}/h.$\cite{wiel} This 
is a highly non-trivial check from the numerical point of view, and proves 
that our finite system approach is capable of sustaining a fully developed 
Kondo peak with the exact spectral weight. Within our approach, the Coulomb 
blockade peaks become discernible through the application of a magnetic field. As 
seen in Fig. 2, with increasing magnetic field, spin-fluctuations at 
the impurity site are progressively quenched, and the associated enhancement 
of the conductance turns into a valley 
flanked by two Coulomb blockade peaks roughly separated by $U$. 
On the other side, the Coulomb blockade peaks are rather insensitive to 
magnetic field effects. 
 
\begin{figure}[tbp] 
\vspace{-.9cm}
\epsfxsize=7.5cm 
\epsfysize=8cm 
\centerline{\epsfbox{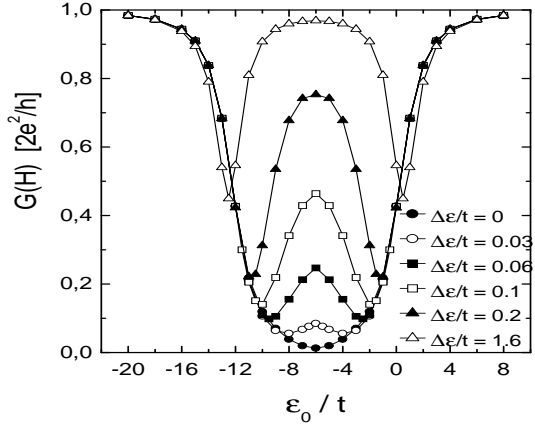}} 
\narrowtext 
\vspace{-1.5cm}
\caption{Same as Fig. 2 for the {\it side} dot configuration ($t_i/t=1$).} 
\end{figure} 
 
The equivalent to the traditional Kondo effect, corresponding to the 
geometrical arrangement of Fig. 1b, is shown in Fig. 3. In this 
configuration, the conductance reaches the unitary limit either when the dot 
is fully occupied ($\varepsilon _{0}+U<0$) or empty ($%
\varepsilon _{0}>0$). In both cases, the {\it side} 
dot weakly perturbs the transmission along the tight-binding chain, as the 
possible scattering processes disappear. On the other side, the conductance 
becomes progressively blocked as the {\it side} dot enters in the Kondo 
regime, reaching the anti-unitary limit $G(0)=0$ exactly at the {\it side} 
dot symmetric configuration $\varepsilon _{0}=-U/2.$ In other words, even 
though the non-interacting central site provides, in principle, a channel 
for transmission, through its coupling to the {\it side} dot it becomes a 
perfectly reflecting barrier. 
 
The results for the total ($\rho _{\uparrow }+\rho _{\downarrow }$) local 
density of states (DOS) shown in Fig.(4) provides a nice qualitative 
explanation of the linear conductance results in both geometrical 
arrangements. Figs. (4a) and (4b) correspond to the local DOS at sites 0 and at the dot 
respectively, both for the {\it side} dot configuration. Starting with Fig. (4b), a 
well-defined Kondo resonance is discernible around the Fermi level in 
absence of magnetic field. The exact (unitary) zero-field result $%
\rho _{imp}(\omega=0)=t/(\pi t^{\prime \text{ }%
2})=2/(\pi t)\simeq 0.64$ (for $t=1$ and $t'/t=1/\sqrt{2}$) is recovered from our numerical 
approach, as discussed above. 
 
For these parameters, the local DOS at the impurity site in the {\it side} dot configuration is 
equivalent to the local DOS at the impurity in the {\it substitutional} dot 
configuration, and gives rise to the unitary limit $G(0)=2e^{2}/h$ discussed 
above. From the full-width of the zero-field impurity DOS at half-maximum 
(FWHM), we estimate $k_{B}T_{K}/t\simeq 0.3$ for these parameters; 
note that this estimate agrees qualitatively with the magnetic field values 
for which the Kondo effect is destroyed (see Fig. 3). In the presence of a magnetic field, the 
Kondo resonance splits into 
two peaks, generating a local minimum between them. As the conductance in 
the {\it substitutional} dot configuration is proportional to the dot DOS at 
the Fermi level, this explains the abrupt decrease of $G(H)$ with increasing 
$H$ at the middle of the Kondo valley shown in Fig. 2. 
 
\begin{figure} 
\vspace{-.5cm}
\epsfxsize=9.5cm 
\epsfysize=10cm 
\centerline{\epsfbox{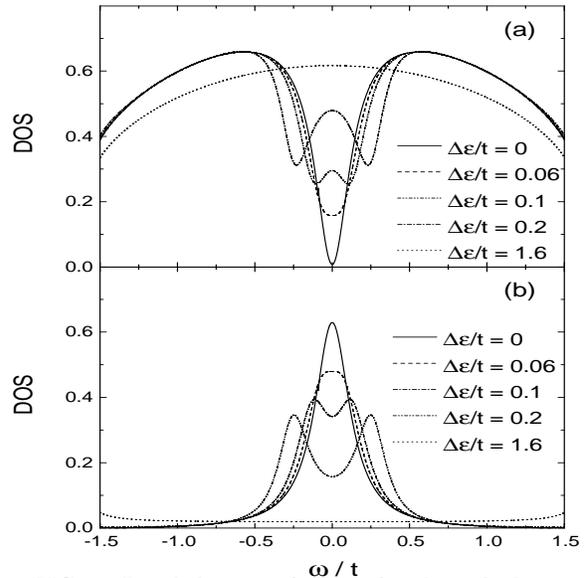}} 
\narrowtext 
\vspace{-1.2cm}
\caption{Local density of states for the {\it side} dot configuration 
 corresponding to the symmetric situation $\varepsilon_0 =-U/2$ for different 
values 
 of magnetic 
field: a) at the 
 non-interacting site 0 of the tight-binding chain; b) at the QD site (same 
parameters as in 
Fig. 3).} 
\end{figure} 
 
Coming back to the {\it side} dot, the most noticeable 
features of Fig. (4a), corresponding to the DOS at site 0 of the chain, are the 
profound dips it exhibits around the Fermi level; a pseudo-gap appears for $H=0$ at the 
symmetric situation $\varepsilon _{0}=-U/2$ exactly at the Fermi level. The 
existence of this pseudo-gap explains the conductance minimum of Fig. 3 at $%
\varepsilon _{0}=-U/2.$ In the presence of a magnetic field, the dip weakens 
and accordingly the conductance starts to increase. After a certain 
threshold field, the DOS develops a double-well shape around the Fermi 
level; the distance between the two well minima is about $2\Delta 
\varepsilon$. \cite{refer} 
If the magnetic field is strong enough $(\Delta \varepsilon 
/k_{B}T_{K}\gg 1)$, the Kondo effect is destroyed, the associated coupling 
between the quantum wire and the side dot essentially vanishes, and the DOS 
at site $0$ recovers the semi-elliptical shape corresponding to the 
non-interacting chain with $t=\sqrt{2}t'=t_i$. 
 
Conceptually, the simplest way to understand these transport features is 
using the framework developed by Fano forty years ago.\cite{fano} He 
analyzed the properties of a system consisting of a continuous spectrum 
degenerated with a discrete level, both non-interacting. Under these 
conditions, a dip develops in the density of states of the continuous 
spectrum, as a result of its interaction with the discrete level. In our case, 
the continuous spectrum is provided by the tight-binding chain, while the 
role of the discrete level is played by the Kondo peak at the DOS of the 
{\it side} dot. The Kondo peak in the local DOS at the {\it side} dot is produced at the expense 
of a decrease of the local DOS at the neighbouring site 0. 
As the Kondo effect is destroyed by the magnetic field, 
its associated many-body Fano antiresonance weakens. An 
interesting feature of our calculation is the evolution of $G(H)$ for 
increasing $H$: as the Kondo effect is destroyed, the wide minimum develops 
a high conductance region around the Fermi level. At high fields, the 
conductance shows two dips, roughly separated by $U.$ These two dips are 
again quite natural in the Fano framework: the {\it side} dot DOS, besides 
the Kondo peak, has two single-particle resonances at $\varepsilon _{0}$ 
and $\varepsilon _{0}+U$ (not shown in Fig. 4b)$.$ They give rise to the 
Coulomb blockade peaks in the {\it substitutional} dot configuration (see 
Fig. 2 at a high magnetic field situation). However, in the {\it side} dot 
configuration, they play the role of two discrete levels which also 
produce a Fano antiresonance when these levels coincide with the Fermi energy 
($\varepsilon _{0}=0$ and $\varepsilon _{0}+U=0$): the result 
is now two Coulomb {\it dips} instead two Coulomb blockade {\it peaks}. In 
view of this analysis, it is clear that the wide valley of Fig. 3 results 
from the superposition of two effects: one due to the Kondo effect  around  $\varepsilon 
_{0}=-U/2$, and the other due to charge fluctuations between the dot and 
the leads around $\varepsilon_{0}=0$ and $\varepsilon _{0}+U=0.$ 
 
We believe that our calculations shed light on recent experiments by 
G\"{o}res {\it et al.}\cite{gores}{\it \ }. Using the same samples of Ref. (1), 
and changing the transmission of the left and right tunnel barriers which 
connect the dot to the conducting leads, they perform conductance 
measurements in the Fano regime (strong coupling leads-dot), the Kondo 
regime (intermediate coupling), and the Coulomb blockade regime (weak 
coupling leads-dot). The main results concern the Fano regime, where the 
conductance shows asymmetric Fano dips on top of a slowly varying 
background. Some features of the dips are reminiscent of the Kondo effect, 
as the temperature dependence of their amplitude (Fig. 4 in Ref. 10), and others of the Coulomb 
blockade effect, as the typical 
diamond-shaped structure in differential conductance measurements (Fig. 5 in 
Ref. 10). Both features are easily explained by our results. 
While our calculation is valid for $T=0,$ 
it seems reasonable to expect similar qualitative behavior by increasing the 
temperature at $H=0$: a wide valley for $T<T_{K}$ and two narrow dips separated by $U$ for $%
T>T_{K}.$ Accordingly, 
one can expect a strong dependence of the dip 
amplitude with temperature, decreasing for increasing 
temperature. Besides, the behavior of dips in differential conductance 
measurements should be completely analogous to the related Coulomb blockade 
peaks, leading to diamond-shaped structures for the dips. Our results also 
shed light on the somehow related problem of the persistent current in a 
mesoscopic ring with a {\it side} dot.\cite{affleck,eckle} The results 
remain controversial on this issue, as Ref. 21 found a detrimental 
effect of the {\it side} dot on the persistent current when the Kondo effect 
is operative. An opposite result was found in Ref. 22, with the 
ring exhibiting a perfect (unitary) persistent current in the Kondo regime. 
Our results for the open configuration of the present work provide naturally 
strong support to the detrimental effect found by Affleck and Simon.\cite 
{affleck} 
 
The authors are grateful to C. A. Balseiro for several illuminating 
discussions. The authors acknowledge support from CONICET. 
This work was done partially under grants PIP 
0473/98 (CONICET), PICT97 03-00121-02152 (ANPCyT) and PICT 03-03833 (ANPCyT).


\begin{references}
\bibitem{wiel}  It is interesting to note that only recently this unitary 
limit has been experimentally achieved in Ref. 7.
\end{references}
\end{document}